\documentclass{article}
\usepackage{amssymb}
\usepackage{amsmath,bm}

\title{Theoretical Aspects of the Equivalence Principle}
\author{Thibault Damour}
\date{\it Institut des Hautes \'Etudes Scientifiques, 35, route de Chartres, 91440 Bures-sur-Yvette, France}

\begin{document}

\maketitle

\begin{abstract}
We review several theoretical aspects of the Equivalence Principle (EP). We emphasize the unsatisfactory fact that the EP maintains the absolute character of the coupling constants of physics while General Relativity, and its generalizations (Kaluza-Klein, $\ldots$, String Theory), suggest that all absolute structures should be replaced by dynamical entities. We discuss the EP-violation phenomenology of dilaton-like models, which is likely to be dominated by the linear superposition of two effects: a signal proportional to the nuclear Coulomb energy, related to the variation of the fine-structure constant, and a signal proportional to the surface nuclear binding energy, related to the variation of the light quark masses. We recall the various theoretical arguments (including a recently proposed anthropic argument) suggesting that the EP be violated at a small, but not unmeasurably small level. This motivates the need for improved tests of the EP. These tests are probing new territories in physics that are related to deep, and mysterious, issues in fundamental physics.
\end{abstract}

\section{Introduction}

The Equivalence Principle (EP) is at the heart of the theory of General Relativity (GR). However, the EP should not be counted among the basic principles of physics (such as the principle of conservation of energy, or the least action principle). Historically, the EP is just a heuristic generalization of the experimental fact that all (neutral) bodies seem to fall with the same acceleration in an external gravitational field. Actually, Einstein initially called it the ``hypothesis of equivalence'' (between gravitation and inertia). He elevated its name to that of the {\it principle} of equivalence only later, after he realized how useful this idea was for building a new theory of gravitation generalizing the theory of Special Relativity in a natural manner. Let us recall that, a posteriori, i.e. after the construction of GR, the EP is a consequence of one of the two basic postulates of GR, namely the postulate of a {\it universal coupling} between matter and gravity, obtained by replacing the Poincar\'e-Minkowski metric $\eta_{\mu\nu}$ entering the special relativistic laws of Nature by a curved spacetime metric $g_{\mu\nu} (x)$. In other words, this first postulate says that the ``matter'' part of the total action reads
\begin{equation}
\label{eq1.1}
S_{\rm matter} = S_{\rm SM} [\psi , A , \Phi ; g_{\mu\nu} ; g_a , Y , \lambda , \mu] \, ,
\end{equation}
where $\psi$ (Fermions), $A$ (gauge fields) and $\Phi$ (Brout-Englert-Higgs scalar field) are the fields of the usual (special relativistic) Standard Model (SM) of particle physics; where $g_{\mu\nu} (x)$ has replaced $\eta_{\mu\nu}$; and where $g_a$ ($U(1) \times SU(2) \times SU(3)$ gauge couplings), $Y$ (Yukawa couplings), $\lambda , \mu$ (quartic coupling and mass parameter of $\Phi$) are the usual (spacetime independent) parameters entering the SM. The second postulate defining GR concerns the dynamics of the (geometrico-) gravitational field $g_{\mu\nu} (x)$. It says that the total action of the matter-gravity system is the sum
\begin{equation}
\label{eq1.2}
S [\psi , A , \Phi , g_{\mu\nu}] = S_{\rm SM} [\psi , A , \Phi ; g_{\mu\nu} ; g_a , Y , \lambda , \mu] + S_g [g_{\mu\nu} ; G] \, ,
\end{equation}
where
\begin{equation}
\label{eq1.3}
S_g [g_{\mu\nu} ; G] = \int d^4 x \, \sqrt g \, \frac{R(g_{\mu\nu})}{16\pi G} \, ,
\end{equation}
with $G$ denoting Newton's constant of gravitation.

\smallskip

The main aim of the present contribution will be to emphasize some of the unsatisfactory theoretical aspects of the structure just recalled of GR, and to discuss the reasons for expecting that the EP be (apparently) violated, thereby opening an interesting new window on physics beyond the current standard SM $+$ GR description encapsulated in the action (\ref{eq1.2}).

\section{Absolute structures versus dynamical entities}
\setcounter{equation}{0}

GR has deeply transformed one aspect of the general framework of physics. Before 1915, the description of all the laws of physics was based on {\it absolute structures}, i.e. a-priori-given structures, independent of the material content of the universe. Among these absolute (or ``rigid'') structures of pre-GR physics, one should count not only the spacetime structure (either Newton's absolute space and absolute time, or Special Relativity's absolute space-time), but also all the dimensionless coupling constants entering the laws of Nature, such as
\begin{equation}
\label{eq2.1}
\alpha_{\rm EM} = \frac{e^2}{\hbar c} \simeq \frac{1}{137.035 \, 999 \, 7} \, ,
\end{equation}
\begin{equation}
\label{eq2.2}
\frac{m_p}{m_e} \simeq 1836.152 \, 672 \, ,
\end{equation}
\begin{equation}
\label{eq2.3}
\frac{G \, m_e \, m_p}{\hbar c} \simeq 3.216 \times 10^{-42} \, .
\end{equation}
When faced with the specific values of these dimensionless coupling constants, one is naturally led to ask what has determined them. According to Leibniz, one of the basic principles of rational thinking is the {\it Principle of Reason}: ``Nihil est sine ratione'', ``Nothing is without a reason'' (see \cite{Heidegger} for an interesting discussion). What could be the ``reasons'' behind the specific numbers quoted in Eqs.~(\ref{eq2.1})--(\ref{eq2.3})? We do not have any firm answer to this question, but the point I wish to make is that many developments of 20$^{\rm th}$ century physics suggests that there might be some underlying dynamics that determine the values (\ref{eq2.1})--(\ref{eq2.3}), because one should not expect physics to contain any a-priori-given, absolute structures. Let us note in passing that, here, we shall only explore the consequences of this idea at the level of the action principle (\ref{eq1.2}). It would be interesting to apply this {\it Principle of Absence of Absolute Structures} to Quantum Mechanics, which is based on assuming that quantum reality ``lives'' in some given ``rigid'' Hilbert space, endowed with a ``flat'' Hermitian metric $\langle \psi \mid \psi \rangle = \eta_{\bar i j} \, \bar\psi^{\bar i} \, \psi^j$. For instance, one might think, by analogy with Einstein's generalization of Special Relativity $(\eta_{\mu\nu} \to g_{\mu\nu} (x))$, of studying a generalization of Quantum Mechanics based on a ``curved'' version of a Hilbert-space $(\eta_{\bar i j} \to g_{\bar i j} (\psi))$.

\smallskip

Let us briefly review the developments suggesting that various building blocks of physics that were traditionally viewed as absolute (or ``rigid'') must be replaced by (``elastic'') dynamical identities. Soon after Einstein suggested to replace the absolute spacetime framework by the dynamical entity $g_{\mu\nu} (x)$, Kaluza, Klein, Pauli, et al \cite{KKbook} suggested, in essence, to replace the God-given gauge coupling constants $g_1 , g_2 , g_3$ (respectively corresponding to the three factor groups $U(1) \times SU(2) \times SU(3)$) by dynamical fields, associated to the curved geometry of some compactified, extra dimensions. For instance, in the simplest, original Kaluza-Klein case where one considers a five-dimensional version of GR, with the extra (spatial) fifth coordinate $x^5$ compactified in a circle $(0 \leq x^5 \leq 2\pi)$, the off-diagonal components of the metric define a $U(1)$ gauge field $A_{\mu} (x) \sim g_{5\mu} (x) / g_{55} (x)$, and the fine-structure constant $\alpha_1 \sim g_1^2 / \hbar c$ measuring the coupling strength of this $U(1)$ gauge field is proportional to the inverse square of the size of the fifth dimension, i.e.
\begin{equation}
\label{eq2.4}
\alpha_1 \propto g_{55}^{-1} (x) \, .
\end{equation}

Later work (by Klein, Pauli, DeWitt, Kerner,$\ldots$) extended the Kaluza-Klein mechanism to non-abelian gauge groups, such as $SU(2)$ or $SU(3)$. This leads to a generalization of Eq.~(\ref{eq2.4}), with a relation between the corresponding gauge couplings $\alpha_2 \sim g_2^2 / \hbar c$, $\alpha_3 \sim g_3^3 / \hbar c$ and the size of the compactified manifold \cite{KKbook,Kerner:1988fn}.

\smallskip

Another important development of 20$^{\rm th}$ century physics shares with the Kaluza-Klein mechanism the idea of replacing an absolute structure by a dynamical entity: it is the mechanism of dynamical symmetry breaking, and particularly the Brout-Englert-Higgs \cite{Englert:1964et,Higgs:1964pj} mechanism. The implementation (by Weinberg \cite{Weinberg:1967tq},~$\ldots$) of this mechanism in the Standard Model of particle physics has the striking consequence that the masses of the leptons, and notably the mass, $m_e$, of the electron, are no longer absolute quantities that must be a priori put by hand, but become proportional to the vacuum expectation value (VEV), $\langle \Phi \rangle$, of (one component of) the Brout-Englert-Higgs field $\Phi$. Essentially, one finds that, e.g., the electron mass is given by
\begin{equation}
\label{eq2.5}
m_e \sim Y_e \langle \Phi \rangle \, ,
\end{equation}
where $Y_e$ is a {\it dimensionless} Yukawa coupling constant, and where the value $\langle \Phi \rangle$ of the VEV of $\Phi$ is a function of other dynamical parameters, determined by minimizing the $\Phi$-potential $V(\Phi) = -\frac{1}{2} \, \mu^2 \vert \Phi \vert^2 + \frac{1}{4} \, \lambda \vert \Phi \vert^4$, namely $\langle \Phi \rangle = \mu / \sqrt\lambda$.

\smallskip

In the three cases mentioned so far (GR; Kaluza-Klein; dynamical symmetry breaking), the constants appearing in the description of local physics $(\eta_{\mu\nu} ; \alpha_1 , \ldots , m_e , \ldots )$ are no longer God-given structures, but are determined by external dynamical effects: 
\begin{equation}
\label{eq2.n}
\eta_{\mu\nu} = g_{\mu\nu} (x) \ ; \quad \alpha_1 \sim g_{55}^{-1} (x) \ ; \quad m_e \sim Y_e \langle \Phi (x) \rangle \, .
\end{equation}
Note that the last result transfers the task of dynamically fixing $m_e$ to the task of dynamically fixing $\Phi (x)$ by some God-given potential $V(\Phi)$, and has still to a priori assume some dimensionless couplings $(Y_e , \lambda)$, together with some mass scale $\mu$. A more ambitious attempt to dynamically determine {\it all} mass scales in particle physics was developed in the 1980's under the name of no-scale supergravity (see \cite{Lahanas:1986uc}). In this model, all the low-energy mass scales, including those (such as the lepton masses) that are related to a symmetry breaking mechanism, are related to the Planck mass scale
\begin{equation}
\label{eq2.6}
M_P \equiv \sqrt{\frac{\hbar c}{G}} \simeq 2.177 \times 10^{-5} {\rm g}
\end{equation}
via, essentially, exponentially small factors involving various (dimensionless) gauge or Yukawa coupling constants at some high unification scale $\Lambda_{\rm GUT} \sim M_P$. Schematically, this leads to symmetry-breaking scales, and corresponding particle masses, that depend on dimensionless coupling constants by relations of the type
\begin{equation}
\label{eq2.7}
m_i [\alpha] \sim M_P \, e^{-\frac{C_i}{\alpha}} \, ,
\end{equation}
where $\alpha \sim g^2 / \hbar c$ is evaluated at the unification scale, and where $C_i$ are some (calculable) constants of order unity. The same type of relation applies to the proton mass $m_p$ because of the (related) phenomenon of dimensional transmutation \cite{Gross:1973id,Politzer:1973fx}. [Both mechanisms are rooted in the logarithmic nature of the Renormalization Group.]

\smallskip

In addition, within the even more ambitious framework of (Super-)String theory \cite{Becker:2007zj} one expects that all the mechanisms we have been mentioning in Eqs.~(\ref{eq2.4}) --(\ref{eq2.6}) take place simultaneously, so that, finally, none of the coupling constants and masses entering low-energy physics are absolute quantities, but they are all functions of some dynamical fields (called {\it moduli} fields) that measure the size, shape and other physico-mathematical features of the six (or seven) extra dimensions of space that are expected to be compactified down to unobservably small proportions. In the lowest approximation (tree-level approximation) of this generalized Kaluza-Klein mechanism, (some of) the moduli fields are massless (like the graviton and the original Kaluza-Klein scalar $g_{55} (x)$), so that one ends up expecting that the masses and couplings of low-energy physics are function of one or several long-range moduli field $\varphi_A (x)$, say
\begin{eqnarray}
\label{eq2.8}
m_i &= &m_i [\varphi_A (x)] \, , \nonumber \\
\alpha_a &= &\alpha_a [\varphi_A (x)] \, .
\end{eqnarray}

Summarizing: the evolution of (theoretical) physics suggests the validity of a {\it Principle of Absence of Absolute Structures}. From the (retrospective) point of view of this principle, the Equivalence Principle on which GR is based looks quite {\it asymmetric}: it replaces the absolute structure $\eta_{\mu\nu}$ of the special relativistic spacetime by the dynamical field $g_{\mu\nu} (x)$, but it maintains the absolute character of all the other structures entering the action (\ref{eq1.2}): the masses and the various couplings $(g_a , Y , \lambda , \ldots , G)$

\section{Dynamical coupling constants and EP violations}
\setcounter{equation}{0}

The {\it asymmetry} introduced by GR between a soft, dynamical spacetime structure, and a rigid, non-dynamical set of coupling constants (including mass ratios such as $m_p / m_e \simeq 1836$) was questioned, long before the particle physics developments mentioned above, by Dirac \cite{Dirac:1937ti} and Jordan \cite{Jordan}. Dirac phenomenologically assumed that the small dimensionless coupling $G \, m_e \, m_p / \hbar c$ of Eq.~(\ref{eq2.3}) varied proportionally to the inverse of the age of the universe, while Jordan (reviving generalizations of General Relativity \`a la Kaluza-Klein) essentially assumed that both $\alpha_{\rm EM}$ and $G$ could become spacetime fields $\varphi (t,{\bm x})$. Actually, the first author to clearly realize that Jordan's original theory implied that the fine-structure constant $\alpha_{\rm EM}$ had become replaced by a field $\varphi (t,{\bm x})$ was Fierz \cite{Fierz}. Fierz then pointed out that astronomical data (line spectra of galaxies) were putting rather strong constraints on the spacetime variability of $\alpha_{\rm EM}$, and suggested to restrict the original, two-parameter class of Jordan's ``varying constant'' theories to the special one-parameter class where the fine-structure constant $\alpha_{\rm EM}$ remains constant, but where the gravitational coupling $G$ is allowed to become a spacetime field. [This EP-respecting one-parameter Jordan-Fierz theory coincides with the tensor-scalar theory later studied by Brans and Dicke.]

\smallskip

The considerations of Jordan and Fierz on field-theory models of varying constants attracted the attention of Dicke. In particular, Dicke realized the important fact that any theory in which the local coupling constants are spatially dependent will entail some violation of the (weak) Equivalence Principle (EP), namely some non-universality in the free-fall acceleration of bodies embedded in an external gravitational field. Dicke's general argument \cite{Dicke} is that the mass $m_i$ of a body, which is made (in view of $mc^2 = E_{\rm tot} = \sum_{\alpha} \, E_{\alpha}$) of many contributions, related to various interaction energies (strong, weak, electromagnetic; to which we can now add the Brout-Englert-Higgs interactions, responsible for the ``rest masses'' of the quarks and the leptons), is a certain, complicated function of various coupling constants, notably the gauge and Yukawa coupling constants: $m_i = m_i [\alpha_{\rm EM} , \ldots]$. If the coupling constants are spatially dependent, the free-fall acceleration deduced from the action of a point particle embedded in a (general relativistic) gravitational field $g_{\mu\nu} (x)$,
\begin{equation}
\label{eq3.1}
S_{m_i} = - \int m_i [\alpha_{\rm EM} (x) , \ldots] \sqrt{-g_{\mu\nu} (x) \, dx^{\mu} \, dx^{\nu}} \, ,
\end{equation}
will read (in the slow-velocity limit)
\begin{eqnarray}
\label{eq3.2}
{\bm a}_i &= &{\bm g} - {\bm \nabla} \ln m_i [\alpha_{\rm EM} (x) , \ldots] \nonumber \\
&= &{\bm g} - \frac{\partial \ln m_i [\alpha_{\rm EM} , \ldots]}{\partial \, \alpha_{\rm EM}} \, {\bm \nabla} \, \alpha_{\rm EM} - \ldots \, .
\end{eqnarray}
The coefficients associated to the spatial gradients of the various coupling constants in Eq.~(\ref{eq3.2}) are expected not to be universal, so that ${\bm a}_i \ne {\bm a}_j$ if the composition of body $i$ differs from that of body $j$.

\smallskip

To turn the result (\ref{eq3.2}) into an explicit prediction for the composition dependence of the EP-violation parameter
\begin{equation}
\label{eq3.3}
\eta_{ij} \equiv \frac{a_i - a_j}{\langle a \rangle}
\end{equation}
(where the time-honoured notation $\eta_{ij}$ for the E\"otv\"os parameter should not be confused with the above-used notations $\eta_{\mu\nu}$ and $\eta_{\bar i j}$) one needs: (i) an explicit, Jordan-type, field model of the spacetime variability of coupling constants (predicting both the dynamics of the field $\varphi (t,{\bm x})$, and the dependence of $\alpha_{\rm EM}$, $m_p / m_e$, $G \, m_e \, m_p$, $\ldots$, on the field $\varphi$), and (ii) an estimate of the dependence of the mass of a body $i$ (say a chunk of Beryllium) on the various coupling constants of particle physics. Concerning the point (i), several models have been considered in the literature: the original (Kaluza-Klein-)Jordan-type scalar field, coupling (in the ``Einstein frame'') {\it only} to the electromagnetic action, and thereby affecting only $\alpha_{\rm EM}$, has been revived by Bekenstein \cite{Bekenstein:1982eu}. The properties of this model have been studied by several authors (e.g. \cite{Sandvik:2001rv}). Other authors have focussed on the more general type of field models suggested by String Theory, i.e. on ``dilaton models'' where a scalar field $\varphi (t,{\bm x})$ monitors, in a correlated manner, the spacetime variability of, essentially, all the coupling constants: gauge couplings, Yukawa couplings, gravitational coupling,$\ldots$. [Here, we simplify the general case considered in Eq.~(\ref{eq2.8}) above where there could be several (massless) moduli fields $\varphi_A (x)$, to the case of a single modulus field, say $\varphi (x)$, that we conventionally call the ``dilaton''.] In these models, because of the complex dependence of mass on the various couplings (point (ii)), the EP-violation parameter (\ref{eq3.3}) has, in general, a complicated dependence on the nuclear composition of bodies $i$ and $j$ (see \cite{Damour:1994zq}). The dependence of the mass $m_i$ on quark masses, via nuclear interactions, is especially difficult to estimate, see \cite{Flambaum:2002wq,Donoghue:2006du,Damour:2007uv,Dent:2008gu}. The complexity of the composition-dependence of the EP-violation $\eta_{ij}$ in dilaton models is a phenomenologically interesting fact which might allow, in principle, to experimentally probe the existence of a long-range dilaton-like field, via EP tests comparing several different pairs of bodies. Correlatively, the predicted general structure of the composition dependence of $\eta_{ij}$ can be used to optimize the choice of materials in EP experiments \cite{DamourBlaser,Damour:1996xt}. More precisely, a study of the EP violations induced in dilaton models \cite{Damour:1994zq,Dent:2008gu,Damour:2010rm,Damour:2010rp} found that the EP violation (\ref{eq3.3}) has the general structure
\begin{equation}
\label{eq3.4}
\eta_{ij} = \left[ c_1 \, \frac{Z(Z-1)}{A^{4/3}} + \frac{c_2}{A^{1/3}} + c_3 \, \frac{A-2Z}{A} + c_4 \, \frac{(A-2Z)^2}{A^2} \right]_{ij}
\end{equation}
where $i,j$ label two different materials (say made of atoms $i$ or $j$), with $A \equiv N + Z$ denoting the nucleon number, $Z$ denoting the atomic number, and $[Q]_{ij} \equiv Q_i - Q_j$.

\smallskip

Let us briefly indicate the physical origin of the various contributions in Eq.~(\ref{eq3.4}). Let us denote by $k_a$ the various dimensionless coupling constants of which the mass $m_i$ of a certain body depends. We can choose
\begin{equation}
\label{eq3.5}
k_0 = \frac{\Lambda_{\rm QCD}}{M_P} \, , \  k_1 = \alpha_{\rm EM} \, , \  k_2 = \frac{\frac{1}{2} (m_d + m_u)}{\Lambda_{\rm QCD}} \, , \ k_3 = \frac{m_d - m_u}{\Lambda_{\rm QCD}} \, , \ k_4 = \frac{m_e}{\Lambda_{\rm QCD}}
\end{equation}
where $\Lambda_{\rm QCD}$ denotes the quantum chromodynamics (QCD) mass scale. Using the fact that the low-energy physics determining the masses of ordinary matter does not involved the Planck mass, we can write
\begin{equation}
\label{eq3.6}
m_i = \Lambda_{\rm QCD} \, \hat m_i [k_1 , k_2 , k_3 , k_4] \, .
\end{equation}

As mentioned in Eq.~(\ref{eq2.8}) above, in dilaton models one expects the ``constants'' $k_a$ of Eq.~(\ref{eq3.5}) to be some functions of the dilaton $\varphi$. Then, from the total action
\begin{equation}
\label{eq3.7}
S = \int \frac{d^4 x \, \sqrt g}{16\pi G} \, (R - 2 (\partial\varphi)^2) - \sum_i \int m_i [k_a (\varphi (y_i))] \, \sqrt{-g_{\mu\nu} (y_i) \, dy_i^{\mu} \, dy_i^{\nu}}
\end{equation}
one gets a coupled set of equations for $g_{\mu\nu} (x)$, $\varphi (x)$ and for the dynamics of the bodies $i$, namely (with $u_i^{\mu} = dy_i^{\mu} / ds_i$)
\begin{eqnarray}
\label{eq3.8}
R_{\mu\nu} &= &2 \, \partial_{\mu} \, \varphi \, \partial_{\nu} \, \varphi + 8\pi G \left(T_{\mu\nu} - \frac{1}{2} \, T g_{\mu\nu} \right) \, , \\
\label{eq3.9}
\Box_g \, \varphi &= &-4\pi G \sigma \, , \\
\label{eq3.10}
\nabla_{u_i} \, u_i^{\mu} &= &\bar a_i^{\mu} \, ,
\end{eqnarray}
where indices are moved by $g_{\mu\nu}$ and where
\begin{eqnarray}
\label{eq3.11}
T^{\mu\nu} (x) &= &\frac{1}{\sqrt{g(x)}} \sum_i \int ds_i \, m_i [\varphi (y_i)] \, u_i^{\mu} \, u_i^{\nu} \, \delta^{(4)} (x-y_i) \nonumber \\
&\equiv &\sum_i T_i^{\mu\nu} (x) \, , \\
\label{eq3.12}
\sigma (x) &= &- \frac{1}{\sqrt{g(x)}} \sum_i \int ds_i \, \frac{\partial m_i}{\partial\varphi} \, \delta^{(4)} (x-y_i) \nonumber \\
&\equiv &\sum_i \alpha_i [\varphi (x)] \, T_i (x) \, , \\
\label{eq3.13}
\bar a_i^{\mu} &= &- \frac{\nabla_{\perp}^{\mu} \, m_i}{m_i} \equiv - \alpha_i [\varphi (y_i)] \, \nabla_{\perp}^{\mu} \varphi (y_i) \, .
\end{eqnarray}
Here, $T \equiv g_{\mu\nu} T^{\mu\nu}$, $\nabla_{\perp}^{\mu} \equiv \nabla^{\mu} + u_i^{\mu} \, u_i^{\nu} \, \nabla_{\nu}$ and the dimensionless quantity
\begin{equation}
\label{eq3.14}
\alpha_i (\varphi) \equiv \frac{\partial \, \ln (m_i (\varphi)/M_P)}{\partial\varphi}
\end{equation}
measures the strength of the coupling of the dilaton to the particles of type $i$.

\smallskip

Solving these coupled equations, one finds that the Newtonian interaction energy between the masses $m_i$ and $m_j$ has the form
\begin{equation}
\label{eq3.15}
V_{\rm int} = -G \, \frac{m_i \, m_j}{r_{ij}} \, (1 + \alpha_i \, \alpha_j)
\end{equation}
where $\alpha_i = \alpha_i (\varphi_0)$, $\varphi_0$ denoting the VEV of $\varphi$, i.e. its background value far away from the considered masses. In terms of the $\alpha_i$'s, the EP-violation $\eta_{ij}$, Eq.~(\ref{eq3.3}), reads
\begin{equation}
\label{eq3.16}
\eta_{ij} = \frac{(\alpha_i - \alpha_j) \, \alpha_E}{1+\frac{1}{2} (\alpha_i + \alpha_j) \, \alpha_E} \simeq (\alpha_i - \alpha_j) \, \alpha_E \, ,
\end{equation}
when comparing the fall of bodies $i$ and $j$ in the gravitational field generated by the external body $E$.

\smallskip

Using the fact that, in view of Eq.~(\ref{eq3.6}), the $\varphi$-dependence of $m_i/M_P$ enters through the $\varphi$-dependence of $k_0 = \Lambda_{\rm QCD} /M_P$, and $k_1 , k_2 , k_3 ,k_4$, we can write $\alpha_i$, Eq.~(\ref{eq3.14}), as
\begin{eqnarray}
\label{eq3.17}
\alpha_i &= &\sum_a \ \frac{\partial \ln (k_0 (\varphi) \, \hat m_i [k_1 (\varphi) , \ldots , k_4 (\varphi)])}{\partial\varphi} \nonumber \\
&= &d_{k_0} + \sum_{a \ne 0} \ d_{k_a} \, Q_i^{k_a}
\end{eqnarray}
where (for $a=0,1,2,3,4$)
\begin{equation}
\label{eq3.18}
d_{k_a} = \partial \ln k_a (\varphi) / \partial\varphi \, ,
\end{equation}
and where (for $a \ne 0$),
\begin{equation}
\label{eq3.19}
Q_i^{k_a} \equiv \frac{\partial \ln \hat m_i}{\partial \ln k_a} \, .
\end{equation}
Eq.~(\ref{eq3.17}) exhibits a decomposition of the scalar coupling $\alpha_i$ into a sum of factorized contributions: each contribution, $d_{k_a} \, Q_i^{k_a}$, is the product of a model-dependent, fundamental dilaton coupling parameter $d_{k_a}$, Eq.~(\ref{eq3.18}) (which measures how the dilaton modifies the coupling parameters $\Lambda_{\rm QCD} / M_P$, $\alpha_{\rm EM}$, $m_{\rm quark} / \Lambda_{\rm QCD} , \ldots$ entering low-energy physics), and of a phenomenological effective charge, $Q_i^{k_a}$ (which measures how the mass ratio $m_i / \Lambda_{\rm QCD}$ depends on the coupling parameters $\alpha_{\rm EM}$, $m_{\rm quark} / \Lambda_{\rm QCD} , \ldots$). Note that the effective charge associated to $k_0 = \Lambda_{\rm QCD} / M_P$ is simply $Q_i^{k_0} \equiv 1$ because of the factorization $m_i / M_P \equiv (\Lambda_{\rm QCD} / M_P) \, \hat m_i [k_1 , \ldots , k_4]$. The interest of this decomposition is that one can compute the effective charges $Q_i^{k_a}$ independently of any particular dilaton model. For instance, the charge $Q_i^{k_1} = \partial \ln \hat m_i / \partial \ln \alpha_{\rm EM}$ measures the fractional part of the mass $m_i$ which comes from electromagnetic (Coulomb) effects. Ref.~\cite{Damour:2010rp} found the following result for this electromagnetic-coupling effective charge
\begin{equation}
\label{eq3.20}
Q_i^{k_1} = F_i \left[ -1.4 + 8.2 \, \frac{Z}{A} + 7.7 \, \frac{Z(Z-1)}{A^{\frac{4}{3}}} \right]_i \times 10^{-4}
\end{equation}
where $F_i \equiv A_i \, m_{\rm amu} / m_i$, with $m_{\rm amu} = 931{\rm Me} V$ denoting the atomic mass unit (i.e. the nucleon mass $m_N = 939 {\rm Me} V$ minus the average binding energy per nucleon, $\simeq 8 {\rm Me} V$). The factor $F_i$ remains quite close to $1$ all over the periodic table (modulo $O(10^{-3})$). The various contributions on the right-hand side of Eq.~(\ref{eq3.20}) come from electromagnetic effects in the proton mass, in the neutron mass and in the binding energy of the nucleus. The effective charges associated to the light quark masses $m_d$ and $m_u$ (and their symmetric and antisymmetric combinations $k_2$ and $k_3$, Eq.~(\ref{eq3.5})) are more difficult to estimate. Thanks to recent progress \cite{Donoghue:2006du,Damour:2007uv} in the understanding of the quark-mass dependence of nuclear binding, Refs.~\cite{Damour:2010rm,Damour:2010rp} could estimate $Q_i^{k_2}$ and $Q_i^{k_3}$ with some reliability. For instance, the effective charge associated to the variation of the average quark mass $\frac{1}{2} (m_d + m_u)$ was found to be
\begin{equation}
\label{eq3.21}
Q_i^{k_2} = F_i \left[ 0.093 - \frac{0.036}{A^{1/3}} - 0.020 \, \frac{(A-2Z)^2}{A^2} - 1.4 \times 10^{-4} \frac{Z(Z-1)}{A^{4/3}} \right]_i \, .
\end{equation}
Of main phenomenological interest in this effective charge is the term $\propto A^{-1/3}$ which comes from the quark-mass dependence of surface effects in the nuclear binding energy. Let us finally note that the numerically dominant term in the scalar coupling to matter, Eq.~(\ref{eq3.17}), is expected to be the first term, $d_{k_0}$, whose associated charge is simply $Q_i^{k_0} = 1$, independently of the considered body. This comes about because, to leading order (in the chiral limit) the masses of all hadrons are proportional to the QCD mass scale $\Lambda_{QCD}$. Note that this universal contribution cancells out in the difference $\alpha_i - \alpha_j$ entering the EP violation (\ref{eq3.16}). [However, it still plays a role through its presence in $\alpha_E = d_{k_0} + \underset{a \ne 0}{\sum} \, d_{k_a} \, Q_E^{k_a}$.]

\smallskip

Combining the model-independent information about the numerical values of the various charges $Q_i^{k_a}$, with some plausible theoretical expectations about the values of the model-dependent dilaton-couplings $d_{k_a}$, Eq.~(\ref{eq3.18}), it has been argued in Refs.~\cite{Damour:2010rm,Damour:2010rp} that the potentially dominant EP violating effects will be contained in the first two terms on the right-hand side of Eq.~(\ref{eq3.4}): i.e., an EP violation $\propto [Z(Z-1) \, A^{-4/3}]_{ij}$ linked to Coulomb nuclear effects, and an EP violation $\propto [A^{-1/3}]_{ij}$ linked to surface nuclear binding energies. More precisely, Refs.~\cite{Damour:2010rm,Damour:2010rp} suggested that the scalar couplings $\alpha_i$ can be approximately described as the combination of three independent terms:
\begin{equation}
\label{eq3.22}
\alpha_i \simeq d_{k_0}^* + d_{k_1} \, Q'^1_i + d_{k_2} \, Q'^2_i \, ,
\end{equation}
with a simplified Coulomb term (associated to the variation of $k_1 = \alpha_{\rm EM}$)
\begin{equation}
\label{eq3.23}
Q'^1 = 7.7 \times 10^{-4} \, \frac{Z(Z-1)}{A^{4/3}}
\end{equation}
and a simplified surface nuclear energy term (associated to the variation of $k_2 = \frac{1}{2} (m_d + m_u) / \Lambda_{\rm QCD}$)
\begin{equation}
\label{eq3.24}
Q'^2 = - \frac{0.036}{A^{1/3}} - 1.4 \times 10^{-4} \, \frac{Z(Z-1)}{A^{4/3}} \, .
\end{equation}
The numerical variations of these two charges over the periodic table are $Q'^1 ({\rm Pt}) - Q'^1 ({\rm Li}) \simeq 4 \times 10^{-3}$ and $Q'^2 ({\rm Pt}) - Q'^2 ({\rm Li}) \simeq 10^{-2}$. See Table~I for numerical values of these effective charges for a sample of materials.
\begin{table}[h]\centering
\caption{Approximate dominant EP-violating `dilaton charges' for a sample of materials. These charges are averaged over the (isotopic or chemical, for SiO$_2$) composition.}
\begin{tabular}{ccccc}
\\
{\rm Material}&$A$&$Z$&$Q'^1$&$-Q'^2$ \\ \\
{\rm Li}&7&3&0.345 $\times 10^{-3}$&18.88 $\times 10^{-3}$ \\
{\rm Be}&9&4&0.494 $\times 10^{-3}$&17.40 $\times 10^{-3}$ \\
{\rm Al}&27&13&1.48 $\times 10^{-3}$&12.27 $\times 10^{-3}$ \\
{\rm Si}&28.1&14&1.64 $\times 10^{-3}$&12.1 $\times 10^{-3}$ \\
{\rm SiO$_2$}&...&...&1.34 $\times 10^{-3}$&13.39 $\times 10^{-3}$ \\
{\rm Ti}&47.9&22&2.04 $\times 10^{-3}$&10.28 $\times 10^{-3}$ \\
{\rm Fe}&56&26&2.34 $\times 10^{-3}$&9.83 $\times 10^{-3}$ \\
{\rm Cu}&63.6&29&2.46 $\times 10^{-3}$&9.47 $\times 10^{-3}$ \\
{\rm Cs}&133&55&3.37 $\times 10^{-3}$&7.7 $\times 10^{-3}$ \\
{\rm Pt}&195.1&78&4.09 $\times 10^{-3}$&6.95 $\times 10^{-3}$ \\
\end{tabular}
\end{table} 

Finally, one can argue that the EP-violation signal associated to a long-range dilatonlike field should be well described by two parameters, $D_1 = d^*_{k_0} d_{k_1}$ and $D_2 = d^*_{k_0} d_{k_2}$, with
\begin{equation}
\label{eq3.25}
\eta_{ij} \simeq [D_1 \, Q'^1 + D_2 \, Q'^2]_{ij} \, .
\end{equation}
This two-parameter model, with the well specified effective charges (\ref{eq3.23}), (\ref{eq3.24}), can be used as a guideline for comparing and planning EP experiments. Refs.~\cite{Damour:2010rm,Damour:2010rp} used it to perform a joint analysis of the two current experiments which have reached the $10^{-13}$ level in $\eta_{ij}$, namely, the terrestrial E\"otWash experiment \cite{Schlamminger:2007ht} and the celestial Lunar Laser Ranging one \cite{Williams:2004qba}. This resulted in limits of order $10^{-9}$ on the two (unknown) theoretical parameters $D_1$ and $D_2$.

\section{Scenarios for the dynamical selection of the coupling constants}
\setcounter{equation}{0}

Let us come back to the main issue behind our discussion: if one does admit (as suggested in particular by String Theory) that there are no absolute structures in physics and, therefore, in particular, that all the coupling constants $k_a$, Eq.~(\ref{eq3.5}), entering low-energy physics are related to underlying dynamical entities, can one make some observable predictions about the values and variability of coupling constants, and thereby about the level of EP violation? We saw in the previous Section that the general idea of the absence of absolute structures, combined with the knowledge of low-energy physics, did lead to some predictions, such as the generic structure (\ref{eq3.4}) of possible EP violations, and with some minimal assumptions, the probable dominance of only two effective charges in EP signals, see Eq.~(\ref{eq3.25}). However, this phenomenological analysis did not give any clue on the plausible values of the model-dependent parameters $d_{k_a}$, Eq.~(\ref{eq3.17}), and $D_a = d_{k_0}^* \, d_{k_a}$, determining the {\it strength} of the EP violation. In addition, we have been assuming that there existed at least one {\it massless} modulus field, leading to long range effects. Is this assumption plausible, and is it naturally compatible with the existing, very stringent tests of the EP?

\smallskip

Initially, string theorists hoped that the stringent consistency requirements of string theories would somehow select a unique, stable ``vacuum'', in which consistency requirements and energy minimization would oblige the moduli fields $\varphi_A (x)$ determining the coupling constants of low-energy physics to take particular values $\langle \varphi_A (x) \rangle = \varphi_A^0$. This would be a striking vindication of Leibniz's Principle of Reason. So far it has not been possible to uncover such stringent vacuum-selecting consistency requirements. As a substitute to this grand hope of finding a {\it unique} consistent vacuum, many string theorists hope that there exists a {\it ``discretuum''} of consistent string vacua, i.e. a discrete set of vacua, in each of which the moduli fields take particular values $\varphi_A^0$, corresponding to some discrete, local minimum of the total energy (for reviews see \cite{Douglas:2006es,Denef:2008wq}). If that is the case, this would predict that the coupling constants do not have any temporal or spatial variability because, like in the Brout-Englert-Higgs mechanism, a fluctuation $\delta \varphi_A (x) = \varphi_A (x) - \varphi_A^0$ has an energy cost $\delta V (\varphi_A) \simeq \frac{1}{2} (\partial^2 V / \partial \varphi_A^0 \partial \varphi_B^0) \, \delta \varphi_A \, \delta \varphi_B$ which implies that $\delta \varphi_A (x)$ is a massive, short-ranged field (with Yukawa-type, exponentially suppressed effects). Though such a mechanism might entail observable short-range modifications of gravity \cite{Antoniadis:2007uz}, it predicts the absence of any long-range EP violations. Note that, far from providing no motivation for EP tests, the current majority view of string theorists does imply that EP tests are important: indeed, they represent tests of a widespread theoretical assumption, that any EP-violation observation would refute, thereby teaching us a lot about fundamental issues\footnote{I thank Mike Douglas for suggesting this positive way of formulating the potential theoretical impact of EP tests within the current string-theory majority view.}.

\smallskip

On the other hand, as the current attempts at stabilizing all the string-theory moduli fields (see, e.g., \cite{Denef:2008wq}) are extremely complex and look rather unnatural, one cannot help thinking that there might exist other ways in which string theory (or whatever theory reconciles General Relativity with Particle Physics) connects itself with the world as we observe it. In particular, we know that one of the (generalized) ``moduli fields'', namely the Einsteinian gravitational field $g_{\mu\nu} (x)$, plays a crucial role in determining the structure of the particle physics interactions via the fact that, in a local laboratory, one can approximate, to a high accuracy, a spacetime varying $g_{\mu\nu} (x)$ by a constant Poincar\'e-Minkowski metric $\eta_{\mu\nu}$. In other words, when listing the dimensionless coupling constants (\ref{eq2.1})--(\ref{eq2.3}),$\ldots$, of particle physics one should include $\eta_{\mu\nu} = {\rm diag} \, (-1 , +1 , +1 , +1)$ in the list, and remember that it comes from a long-range, cosmologically evolving field $g_{\mu\nu} (x)$. In this connection, let us further recall that the ``dilaton'', $\varphi (x)$, i.e. the moduli field which determines the value of the basic, ten-dimensional string coupling constant $g_s$ can be viewed (\`a la Kaluza-Klein) as an additional metric component $g_{11 \, 11} (x)$, measuring the size of a compactified eleventh dimension \cite{Witten:1995ex}. This family likeness between the dilaton $\varphi (x)$ and the metric $g_{\mu\nu} (x)$ (which entails a correlated likeness, say in heterotic string theory, between $g_{\mu\nu} (x)$ and the gauge couplings $g_a^2 (x)$, as well as the string-frame gravitational coupling $G(x)$) suggests that there might exist consistent string vacua where some of the moduli fields are not stabilized, but retain their long-range, spacetime-dependent character. As recalled above, such a situation would entail long-range violations of the EP. How come such violations have not yet been observed, given the exquisite accuracy of current tests of the universality of free fall (at the $10^{-13}$ level \cite{Schlamminger:2007ht}) and of current tests of the variability of coupling constants \cite{Rosenband:2008}? A possible mechanism for reconciling a long-range, spacetime varying dilaton (or, more generally, moduli) field $\varphi (x)$ with the strong current constraints on the time or space variability of coupling constants is the {\it cosmological attractor} mechanism \cite{Damour:1992kf,Damour:1994zq,Damour:2002mi} (for other attempts at using cosmological dynamics to stabilize the moduli fields see \cite{Greene:2007sa} and references therein). A simple realization of this mechanism is obtained by assuming that all the coupling functions $B_A (\varphi)$ of $\varphi$ to the fields describing the sub-Planckian particle physics (inflaton, gauge fields, Brout-Englert-Higgs field, leptons, quarks,$\ldots$) admit a limit as $\varphi \to +\infty$ (``infinite bare strong coupling'') \cite{Veneziano:2001ah}. Under this very general, technically simple (but physically highly non trivial) assumption, one finds that the inflationary stage of cosmological expansion has the effect of naturally driving $\varphi$ towards values so large that the present observational deviations from General Relativity are compatible with all the current tests of Einstein's theory \cite{Damour:2002mi,Damour:2002nv}. This ``runaway dilaton'' mechanism also yields an interesting connection between the deviations from General Relativity and the amplitude of large-scale cosmological density fluctuations coming out of inflation. In particular, the level of EP violation is predicted to be
\begin{equation}
\label{eq7}
\eta \equiv \frac{\Delta a}{a} \sim 5 \times 10^{-4} \, k \left( \frac{\delta \rho}{\rho} \right)^{\frac{8}{n+2}} \, ,
\end{equation}
where $k = (b_F / (c \, b_{\lambda}))^2$ is a combination of unknown dimensionless parameters expected to be of order unity, and where $\delta \rho / \rho$ denotes the amplitude of large-scale cosmological density fluctuations, while $n$ denotes the exponent of the inflationary potential $V(\chi) \propto \chi^n$. Inserting the value observed in our universe, $\delta \rho / \rho \sim 5 \times 10^{-5}$, and the value $n=2$ corresponding to the simplest chaotic inflationary potential $(V(\chi) = \frac{1}{2} \, m_{\chi}^2 \, \chi^2)$, the rough prediction (\ref{eq7}) yields $\eta \sim k \times 10^{-12}$ which, given that $k$ is only constrained ``to be of order unity'', is compatible with current EP tests. Note that this runaway dilaton mechanism then predicts (if $n=2$) that a modest increase in the accuracy of EP tests might detect a non zero violation. Note also the rationally pleasing aspect (reminiscent of Dirac's large number hypothesis \cite{Dirac:1937ti}) of Eq.~(\ref{eq7}) which connects the level of variability of the coupling constants to cosmological features (see \cite{Damour:2002nv} for further discussion of this aspect), thereby explaining ``why'' it is so small without invoking the presence of unnaturally small dimensionless numbers in the fundamental Lagrangian.

\smallskip

The ``runaway dilaton'' mechanism just mentioned was formulated as a possible way of reconciling, within a string-inspired phenomenological framework, a ``cosmologically running'' massless\footnote{Let us note in passing that an interesting generalization of the cosmological attractor mechanism is obtained by combining the attraction due to the coupling of $\varphi$ (via $B_A (\varphi)$) to the matter density, to the effect of a quintessence-like potential $V(\varphi)$ \cite{Khoury:2003rn}.} dilaton with observational tests of General Relativity. Let us note that some authors \cite{Wetterich:2008sx,Rabinovici:2007hz} have suggested that the puzzle of having an extremely small vacuum energy $\rho_{\rm vac} \lesssim 10^{-123} (m_{\rm Planck})^4$ might be solved by a mechanism of spontaneous breaking of scale invariance of some (unknown) underlying scale-invariant theory. Under the assumption that scale-invariance is re-established only when a certain ``dilaton field''\footnote{Beware that, here, the name ``dilaton field'' refers to a field, say $\phi$, connected to scale invariance. Such a field $\phi$ is, a priori, quite different from the ``dilaton field'' $\varphi$ of string theory. Indeed, string theory, as we currently know it, contains a basic mass (and length) scale, even in the limits $\varphi \to - \infty$ ($m_s^{(D=10)} = 1 / \sqrt{\alpha'}$) or $\varphi \to + \infty$ ($m_{\rm Planck}^{(D=11)} = 1 / \ell_{\rm Planck}^{(D=11)}$).} $\phi \sim \ln \chi \to \infty$, it seems \cite{Wetterich:2008sx} that a ``$\phi$-dilaton runaway'' behaviour (technically similar to the $\varphi$-dilaton runaway mentioned above) might take place and entail similar observational violations of the EP.

\section{Dynamics versus Anthropics}
\setcounter{equation}{0}

We have mentioned above various visions of the ``reason'' behind the selection of the observed values (\ref{eq2.1})--(\ref{eq2.3}),$\ldots$ of the coupling constants. The intellectually most satisfactory one (given the historical pregnancy of the Principle of Reason \cite{Heidegger}) would be the discovery of subtle consistency requirements which would select an essentially unique physico-mathematical scheme describing the only possible physical laws. In this vision, all the dimensionless numbers of Eqs.~(\ref{eq2.1})--(\ref{eq2.3}),$\ldots$ would be uniquely determined. Note that the discovery of asymptotic freedom and dimensional transmutation, see \cite{Gross:1973id,Politzer:1973fx}, has opened the way to a conceivable rational explanation of very small dimensionless numbers, such as Eq.~(\ref{eq2.3}) (which baffled Dirac): they might be exponentially related to smallish coupling constants, along the model $\Lambda_{\rm QCD} / \Lambda \sim \exp (-8\pi b / g^2)$ where $g^2$ is a  gauge coupling constant considered at the (high-energy, cut-off) scale $\Lambda$. As mentioned above, see Eq.~(\ref{eq2.7}), no-scale supergravity is a generalized version of this mechanism.

\smallskip

In absence of precise clues for realizing this vision, we are left with two types of less satisfactory visions. In one, the extremely vast ``landscape of string vacua'' can dynamically channel the coupling constants towards a discretuum of possible ``locally special values''. This leaves, however, open the problem of finding the ``reason'' why our world has selected one particular set of such, energy-minimizing locally special values. In the other, all (or some of\footnote{Indeed, one can evidently mix the two different scenarios.}) the coupling constants are, like the metric of spacetime around us, dynamically determined by some global aspects of our universe. Both visions contain a partial dynamical ``reason'' behind the selection of the coupling constants, but both visions leave also a lot of room to contingency (or environmental influences). Many authors have suggested that a complementary ``reason'' behind the selection of the coupling constants that we observe, might be the (weak) ``Anthropic Principle'', i.e. the tautological requirement that the physical laws and conditions around us must be compatible with the existence of information processing organisms able to wonder ``why'' the world around them is as it is. In other words, this is essentially an issue of Bayesian statistics: one should consider only a posteriori questions, rather than a priori ones. Though the appeal to such an a posteriori consistency requirement is intellectually less thrilling than the demand of a stringent a priori consistency requirement, it might have satisfied Leibniz. Indeed, Leibniz was one of the enthusiastic historical proponents of the {\it ``Principle of Plenitude''} \cite{Lovejoy} which considers that all logically possible ``things'' (be they objects, beings or, even, worlds) have a tendency to (and therefore {\it must}, if one does not want contingency -- be it God's whim --  to reign) {\it exist}. In addition, in spite of its tautological character, the anthropic consistency requirement does lead to some well-defined, and scientifically interesting (as well as challenging) questions. Indeed, the general scientific question it raises is: what would change in the world around us if the values of the coupling constants (\ref{eq2.1})--(\ref{eq2.3}), etc., would be different? In its generality, this is a very difficult question to address. Let us mention here some scientifically interesting partial answers. [For the fascinating issue of what happens when one varies the vacuum energy density (or cosmological constant) see Refs.~\cite{Weinberg:1987dv,Vilenkin:1994ua}, which predicted that one should observe a non zero $\rho_{\rm vac}$ {\it before} any data had solidly suggested it.] The ``Atomic Principle'' refers to the scientific study of the range of coupling parameters compatible with the existence of the periodic table of atoms, as we know it. In particular, one might ask what happens when one changes the ratio $m_q / \Lambda_{\rm QCD}$ of light quark masses (or the Brout-Englert-Higgs vacuum expectation value which monitors the quark masses) to the QCD energy scale. This issue has been particularly studied by Donoghue and collaborators \cite{Agrawal:1997gf,Agrawal:1998xa,Donoghue:2007zz,Donoghue:2009me}. Recent progress \cite{Damour:2007uv} has shown that the existence of heavy atoms is quite sensitive to the light quark masses. If one were to increase the mass ratio $(m_u + m_d) / \Lambda_{\rm QCD}$ by about 40\%, all heavy nuclei would unbind, and the world would not contain any non trivial chemistry.

\smallskip

Coming back to the issue of EP violation, one might use the idea of a (partially) anthropic selection of coupling constants to predict that the Equivalence Principle should be violated at some level \cite{Damour:2010rp}. Indeed, as in the case of the vacuum energy mentioned above, the observed values of the coupling constants (as well as that of their temporal and/or spatial gradients or variability) should only be required to fall within some life-compatible range, and one should not expect that they take any special, more constrained value, except if this is anthropically necessary. When one thinks about it, one can see some reasons why too strong a violation of the universality of free fall might drastically change the world as we know it, but, at the same time, one cannot see any reason why the EP should be rigorously satisfied. Therefore, one should expect to observe
\begin{equation}
\label{eq5.1}
\frac{\Delta a}{a} \sim \eta_* \ne 0 \, ,
\end{equation}
where $\eta_*$ is the maximum value of $\eta \equiv \Delta a / a$ tolerable for life \cite{Damour:2010rp}. It is a challenge to give a precise estimate of $\eta_*$, but the prediction (\ref{eq3.8}) gives an additional motivation for EP tests. Let us emphasize that the problem of determining, or at least of giving an upper bound to, $\eta_*$ is a scientifically rather well-posed problem. For instance, one of the necessary conditions for the existence of life is the existence of solarlike planetary systems stable over billions of years. A sufficiently large $\eta \ne 0$ will jeopardize this stability, notably under the influence of external, passing stars. The current very small level of EP-violation ensures that stars passing at a distance $D$ disturb the inner dynamics of the solar-system only through tidal effects that decrease like $D^{-3}$. An EP-violation $\eta$ would increase this disturbing effect to a level $\propto \eta \, D^{-2}$. It is also a well-posed problem (that could be resolved by running long-term simulations of the solar-system dynamics in presence of some EP-violation) to determine the level $\eta$ which would destabilize the solar-system through internal EP-violating gravitational effects.

\section{Composition-independent versus composition-dependent tests of gravity}
\setcounter{equation}{0}

To complete our brief review of the theoretical aspects of composition-dependent, EP tests, let us mention the fact that dilaton models of the general type of Eq.~(\ref{eq3.7}) provide a framework in which one can simultaneously discuss, and compare, composition-dependent and composition-independent test of gravity. As discussed in Refs.~\cite{Damour:1994zq,Damour:1992kf,Damour:2002mi,Damour:2002nv,Damour:2010rp,Damour:1992we}, the scalar contribution to gravity, i.e. the extra term $\alpha_i \alpha_j$ in the Newtonian interaction potential (\ref{eq3.15}) is directly linked to post-Newtonian modifications of gravity, as measured by a (composition-dependent) Eddington parametrized post-Newtonian parameter $\gamma_{ij} \ne 1$. More precisely,
\begin{equation}
\label{eq6.1}
1 - \gamma_{ij} = \frac{2 \, \alpha_i \alpha_j}{1+\alpha_i \alpha_j} \, .
\end{equation}
However, in dilaton models $\alpha_i$ is of the form (\ref{eq3.17}), (\ref{eq3.22}) where the composition-independent contribution $d_{k_0}^*$ to $\alpha_i$ is expected to numerically dominate over the composition-dependent contributions $d_{k_a} \, Q'^a_i$. [Indeed, we expect $d_{k_a} \lesssim d_{k_0}^*$, while the effective charges $Q'^a_i$ are $\lesssim 10^{-2}$, see Table~I.] This suggests that the Eddington parameter $\gamma_{ij}$ will be approximately composition-independent, $\gamma_{ij} \simeq \gamma$ and given by
\begin{equation}
\label{eq6.2}
1-\gamma \simeq 2(d_{k_0}^*)^2 \, .
\end{equation}

On the other hand, the EP-violation (\ref{eq3.16}) will be
\begin{equation}
\label{eq6.3}
\eta_{ij} \simeq (\alpha_i - \alpha_j) \, d_{k_0}^* \simeq d_{k_0}^* \, d_{k_1} (Q'^1_i - Q'^1_j) + d_{k_0}^* \, d_{k_2} (Q'^2 _i - Q'^2_j) \, .
\end{equation}
Considering for instance a Be Pt pair (for which $Q'^1_i - Q'^1_j \simeq 4 \times 10^{-3}$ and $Q'^2_i - Q'^2_j \simeq 10^{-2}$), and assuming for simplicity that the effect of $Q'^2_i - Q'^2_j$ dominates, we see that we can write the approximate link
\begin{equation}
\label{eq6.4}
\eta_{\rm Be \, Pt} \simeq 10^{-2} \, \frac{d_{k_2}}{d^*_{k_0}} \, \frac{1-\gamma}{2} \, .
\end{equation}

Now, we have 
\begin{equation}
\label{eq6.5}
d_{k_0}^* \simeq d_{k_0} = \frac{\partial \ln (\Lambda_{\rm QCD} / M_P)}{\partial\varphi}
\end{equation}
while, denoting $m_q \equiv \frac{1}{2} (m_d + m_u)$, 
\begin{eqnarray}
\label{eq6.6}
d_{k_2} &= &\frac{\partial \ln (m_q / \Lambda_{\rm QCD})}{\partial \varphi} = \frac{\partial \ln (m_q / M_P)}{\partial \varphi} - \frac{\partial \ln (\Lambda_{\rm QCD} / M_P)}{\partial \varphi} \nonumber \\
&= &\frac{\partial \ln (m_q / M_P)}{\partial \varphi} - d_{k_0} 
\end{eqnarray}

As discussed in \cite{Damour:2010rp} $d_{k_0}$, Eq.~(\ref{eq6.5}), is expected to be of order 40 (because of the logarithmic enhancement factor $\ln (\Lambda_{\rm QCD} / M_P) \sim 40$), while $d_{k_2}$ is the difference of two terms of order 40. Barring a precise cancellation among the two contributions to $d_{k_2}$, we can expect to have $d_{k_2} \sim 40 \sim d_{k_0} \simeq d^*_{k_0}$. This suggests, in view of Eq.~(\ref{eq6.4}), that current EP tests ($\eta_{\rm Be \, Ti} \lesssim 10^{-13}$ \cite{Schlamminger:2007ht}) correspond to post-Newtonian tests at the level $(1-\gamma) / 2 \sim 10^{-11}$, i.e. 6 orders of magnitude below the current best post-Newtonian test, namely, the Cassini limit $(1-\gamma)/2 < 10^{-5}$ \cite{Bertotti:2003rm}.

\smallskip

Dilaton models can also be straightforwardly applied to comparing EP tests to atomic-clock tests of the dependence of coupling constants on the gravitational potential. See Refs.~\cite{Dent:2008gu,Will:,Damour:1997bf,Damour:1999ip,Nordtvedt:2002qe}. Let us also mention that dilaton models give a useful framework for considering the cosmological aspects of EP-violations, i.e. the cosmological variation of the coupling constants $k_a$. See, e.g., \cite{Damour:2002nv} and references therein.

\section{Conclusions}
\setcounter{equation}{0}

Despite its name, the ``Equivalence Principle'' (EP) is not one of the basic principles of physics. There is nothing taboo about having an observable violation of the EP. On the contrary, one can argue (notably on the basis of the central message of Einstein's theory of General Relativity) that the historical tendency of physics is to discard any, a priori given, absolute structure ({\it Principle of Absence of Absolute Structures}). The EP gives to the set of coupling constants (such as $\alpha_{\rm EM} \simeq 1/137.0359997$) the status of such an, a priori given, absolute structure. It is to be expected that this absolute, rigid nature of the coupling constants is only an approximation. Many theoretical extensions of General Relativity (from Kaluza-Klein to String Theory) suggest observable EP violations in the sense that the set of coupling constants become related to spacetime varying fields.

\smallskip

However, there is no firm prediction for the observable level of EP violation. Actually, the current majority view about the ``moduli stabilization'' issue in String Theory is to assume that, in each string vacuum, the coupling constants are fixed by an energy-minimizing mechanism which is generically expected to forbid any long-range violation of the EP. This, however, makes EP tests quite important: indeed, they represent crucial tests of a widespread key assumption of string-theory model building. This exemplifies how EP tests are intimately connected with some of the basic aspects of modern attempts at unifying gravity with particle physics.

\smallskip

Some phenomenological models (inspired by string-theory structures, or attempting to understand the cosmological-constant issue) give examples where the observable EP violations would (without fine-tuning parameters) be just below the currently tested level. In these ``dilaton models'' the composition dependence of EP signals is (probably) dominated by {\it two} signals, one (related to the variability of the fine-structure constant $\alpha_{\rm EM}$) proportional to $Z(Z-1) A^{-4/3}$, and the other one (related to the variability of the quark masses) proportional to $A^{-1/3}$. Such (runaway dilaton) models comprise many different, correlated modifications of Einsteinian gravity ($\Delta a/a \ne 0$, $\dot\alpha_{\rm EM} \ne 0$, $\gamma_{\rm PPN} - 1 \ne 0 , \ldots$), but EP tests stand out as our deepest possible probe of new physics. Anthropic arguments also suggest that the EP is likely to be violated at some (life-tolerable) level. Let us hope that the refined EP tests which are in preparation (such as the Microscope mission \cite{microscope}) will open a window on the mysterious physics behind the selection of the coupling constants observed in our world.

\end{document}